\documentstyle[12pt]{article}
\def\fnote#1#2{\begingroup\def\thefootnote{#1}\footnote{#2}\endgroup}

\begin{document}

\vspace{36pt}

\begin{center}
{\bf The Making of the Standard Model}\\

\vspace{12pt}

Steven Weinberg\fnote{*}{weinberg@physics.utexas.edu}\\
{\em Theory Group, Physics Department, University of Texas, \\Austin, 
TX, 78712}
\end{center}

\vspace{12pt}

\footnotesize
This is the edited text of a talk given at CERN on September 16, 2003, as part of
a celebration of the 30th anniversary of the discovery of neutral currents and the
20th anniversary of the discovery of the $W$ and $Z$ particles.  

\normalsize

\vspace{12pt}

I have been asked to review the history of the formation of the Standard Model.  It is natural to tell this story as a sequence of brilliant ideas and experiments, but here I will also talk about some of the misunderstandings and false starts that went along with this progress, and why some steps were not taken until long after they became possible.  The study of what was not understood by scientists, or was understood wrongly, seems to me often the most interesting part of the history of science.  Anyway, it is an aspect of the Standard Model with which I am very familiar, for as you will see in this talk, I shared in many of these misunderstandings.

I'll begin by taking you back before the Standard Model to the 1950's.  It was a time of frustration and confusion.  The success of quantum electrodynamics in the late 1940s had produced a boom in elementary particle theory, and then the market crashed.  It was realized that the four-fermion theory of weak interactions had infinities that could not be eliminated by the technique of renormalization, which had worked so brilliantly in electrodynamics. The four-fermion theory was perfectly good as a lowest-order approximation, but when you tried to push it to the next order of perturbation theory you encountered unremovable infinities.  The theory of strong interactions had a different problem; there was no difficulty in constructing renormalizable theories of the strong interactions like the original Yukawa theory but, because the strong interactions are strong, perturbation theory was useless, and one could do no practical calculations with these theories.  A deeper problem with our understanding of both the weak and the strong interactions was that there was no rationale for any of these theories. The weak interaction theory was simply cobbled together to fit what experimental data was available, and there was no evidence at all for any particular theory of strong interactions.

There began a period of disillusionment with quantum field theory.  The community of theoretical physicists tended to split into what at the time were sometimes called, by analogy with atomic wave functions, radial and azimuthal physicists.  Radial physicists were concerned with dynamics, particularly the dynamics of the strong interactions.  They had little to say about the weak interactions.  Some of them tried to proceed just on the basis of general principles, using dispersion relations and Regge pole expansions, and they hoped ultimately for a pure S-matrix theory of the strong interactions, completely divorced from quantum field theory.  Weak interactions would somehow take care of themselves later.  Azimuthal physicists were more modest.  They took it as a working rule that there was no point in trying to understand strong interaction dynamics, and instead they studied the one sort of thing that could be used to make predictions without such understanding --- principles of symmetry. 

But there was a great obstacle in the understanding of symmetry principles.  Many symmetry principles were known, and a large fraction of them were only approximate.  That was certainly true of isotopic spin symmetry, which goes back to 1936 [1].  Strangeness conservation was known from the beginning to be violated by the weak interactions [2].  Then in 1956 even the sacred symmetries of space and time, P and PT conservation, were found to be violated by the weak interactions [3], and CP conservation was found in 1964 to be only approximate [4].  The $SU(3)$ symmetry of the ``eightfold way'' discovered in the early 1960s [5] was at best only a fair approximation even for the strong interactions.  This left us with a fundamental question. Many azimuthal physicists had thought that symmetry principles were an expression of the simplicity of nature at its deepest level. So what are you to make of an approximate symmetry principle?  The approximate simplicity of nature? 

During this time of confusion and frustration in the 1950s and 1960s there emerged three good ideas.  These ideas took a long time to mature, but have become fundamental to today's elementary particle physics. I am emphasizing here that it took a long time before we realized what these ideas were good for partly because I want to encourage today's string theorists, who I think also have good ideas that are taking a long time to mature. 

The first of the good ideas that I'll mention is the quark model, proposed in 1964 independently by Gell-Mann and Zweig [6].  The idea that hadrons are made of quarks and antiquarks, used in a naive way, allowed one to make some sense of the growing menu of hadrons.  Also, the naive quark model seemed to get experimental support from an experiment done at SLAC in 1968 under the leadership of Friedman, Kendall, and Taylor [7], which was analogous to the experiment done by Geiger and Marsden in Rutherford's laboratory in 1911.  Geiger and Marsden had found that alpha particles were sometimes scattered by gold atoms at large angles, and Rutherford inferred from this that the mass of the atoms was concentrated in something like a point particle, which became known as the nucleus of the atom. In the same way, the SLAC experiment found that electrons were sometimes scattered from nucleons at large angles, and this was interpreted by Feynman and Bjorken [8] as indicating that the neutron and proton consisted of point particles.  It was natural to identify these ``partons'' with Gell-Mann and Zweig's quarks.  But of course the mystery about all this was why no one ever saw quarks. Why, for example, did oil drop experiments never reveal third integer charges? I remember Dalitz and Lipkin at various conferences showing all the successful predictions of the naive quark model for hadron systematics, while I sat there remaining stubbornly unconvinced, because everyone knew that quarks had been looked for and not found.

The second of the good ideas that were extant in the 1950s and 1960s was the idea of gauge (or local) symmetry.  (Of course electrodynamics was much older, and could have been regarded as based on a $U(1)$ gauge symmetry, but that wasn't the point of view of the theorists who developed quantum electrodynamics in the 1930s.)  Yang and Mills [9] in 1954 constructed a gauge theory based not on the simple one-dimensional group $U(1)$ of electrodynamics, but on a three-dimensional group, the group $SU(2)$ of isotopic spin conservation, in the hope that this would become a theory of the strong interactions.  This was a beautiful theory because the symmetry dictated the form of the interactions.  In particular, because the gauge group was non-Abelian (the ``charges'' do not commute with each other) there was a self-interaction of the gauge bosons, like the self-interactions of gravitons in general relativity. This was just the sort of thing that brings joy to the heart of an elementary particle theorist.  

The quantization of non-Abelian gauge theories was studied by a number of other theorists [10], generally without any idea of applying these theories immediately to known interactions.  Some of these theorists developed the theory of the quantization of Yang--Mills theories as a warm-up exercise for the problem they really wanted to solve, the quantization of general relativity.  It took a few years before physicists began to apply the Yang--Mills idea to the weak interactions.  This was in part because in 1954, as you may recall, the beta decay interactions were known to be a mixture of scalar, tensor, and perhaps pseudoscalar four-fermion interactions.  This was the result of a series of wrong experiments, each one of which as soon as it was discovered to be wrong was replaced by another wrong experiment. It wasn't until 1957--58 that it became generally realized that the weak interactions are in fact a mixture of vector and axial vector interactions [11], of the sort that would be produced by intermediate vector bosons.  

Theories of intermediate vector bosons were then developed by several authors [12], but generally, except for the papers by Bludman in 1958 and by Salam and Ward in 1964, without reference to 
non-Abelian local symmetries.  (For instance, with the exceptions noted, these papers did not include the quadrilinear interactions among vector bosons characteristic of theories with non-Abelian local symmetries.)  I will have more to say about some of these papers later.

From the beginning, the chief obstacle to the application of the Yang--Mills approach to theories of either the weak or the strong interactions was the problem of mass.  Gauge symmetry forbids the gauge bosons from having any mass, and it was supposed that any massless gauge bosons would surely have been detected.  In all the papers of ref. 12 a mass was put in by hand, but this would destroy the rationale for a gauge theory; the local symmetry principle that motivates such theories would be violated by the insertion of a mass.  Obviously also the arbitrary insertion of mass terms makes theories less predictive. Finally, through the work of several authors [13] in the 1960s, it was realized that 
non-Abelian gauge theories with mass terms inserted by hand are non-renormalizable, and therefore in this respect do not represent an advance over the original four-fermion weak interaction. 

The third of the good ideas that I wished to mention was the idea of spontaneously broken symmetry: there might be symmetries of the Lagrangian that are not symmetries of the vacuum.  Physicists came to this idea through two rather different routes. 

The first route was founded on a fundamental misunderstanding.  Remember that for some time there had been a problem of understanding the known approximate symmetries.  Many of us, including myself, were at first under the illusion that if you had an exact symmetry of the field equations of nature which was spontaneously broken then it would appear experimentally as an approximate symmetry. This is quite wrong, but that's what we thought.  (Heisenberg continued to believe this as late as 1975 [14].)  At first this seemed to offer a great hope of understanding the many approximate symmetries, like isotopic spin, the 8-fold way, and so on.  Thus it was regarded as a terrible setback in 1961 when Goldstone announced a theorem [15], proved by Goldstone, Salam and myself [16] the following year, that for every spontaneously broken symmetry there must be a massless spinless particle.  We knew that there were no such massless Goldstone bosons in strong-interaction physics --- they would have been obvious many years before --- so this seemed to close off the opportunities provided by spontaneous symmetry breaking.  Higgs [17] in 1964 was motivated by this disappointment to try to find a way out of the Goldstone theorem.  He recognized that the Goldstone theorem would not apply if the original symmetry was not just a global symmetry like isotopic spin conservation, but a gauge symmetry like the local isotopic spin symmetry of the original Yang--Mills theory.  The Goldstone boson then remains in the theory, but it turns into the helicity-zero part of a gauge boson, which thereby gets a mass.  At about the same time Englert and Brout [18] independently discovered the same phenomenon, but with a different motivation: they hoped to go back to the idea of using the Yang--Mills theory to construct a theory of the strong interactions mediated by massive vector bosons.  This phenomenon had also been noted earlier by Anderson [19], in a non-relativistic context.

The second of the routes to broken symmetry was the study of the currents of the semi-leptonic weak interactions, the vector and axial-vector currents.  In 1958 Goldberger and Treiman [20] gave a derivation of a relation between the pion decay constant, the axial vector coupling constant of beta decay, and the strong coupling constant.  The relation worked better than would be expected from the rather implausible approximations used.  It was in order to explain the success of the 
Goldberger--Treiman relation that several theorists [21] in the following years developed the idea of a partially conserved axial-vector current, that is, an axial-vector current whose divergence was not zero but was proportional to the pion field.  Taken literally, this was a meaningless proposition, because any field operator that had the right quantum numbers, such as the divergence of the 
axial-vector current, can be called the pion field.  Nature does not single out specific operators as the field of this or that particle.  This idea was greatly clarified by Nambu [22] in 1960.  He pointed out that in an ideal world, where the axial-vector current was not partially conserved but exactly conserved, the existence of a non-vanishing nucleon mass and axial vector coupling would require the pion to be a particle of zero mass.  At sufficiently small momentum transfer this massless pion would dominate the pseudoscalar part of the one-nucleon matrix element of the axial vector current, which leads to the same Goldberger--Treiman result that had previously motivated the notion of partial current conservation.  Nambu and Jona-Lasinio [23] worked out a dynamical model in which the 
axial--vector current would be exactly conserved, and showed that the spectrum of bound states did indeed include a massless pion. 

In this work there was little discussion of spontaneously broken symmetry.  In particular, because the work of Nambu and his collaborators [24] on soft-pion interactions only involved a single soft pion, it was not necessary to identify a particular broken symmetry group.  In much of their work it was taken to be a simple $U(1)$ symmetry group.  Nambu {\em et al.} like Gell-Mann {\em et al.} [21] emphasized the properties of the currents of beta decay rather than broken symmetry.  Nambu, especially in the paper with Jona-Lasinio, described what he was doing as an analog to the successful theory of superconductivity of Bardeen, Cooper and Schrieffer [25].  A superconductor is nothing but a place where electromagnetic gauge invariance is spontaneously broken, but you will not find that statement or any mention of spontaneously broken symmetry anywhere in the classic BCS paper.  Anderson [19] did realize the importance of spontaneous symmetry breaking in the theory of superconductivity, but he was almost the only condensed matter physicist who did. 

The currents of the semi-leptonic weak interactions remained the preoccupation of Gell-Mann and others, who proposed working with them the way Heisenberg had worked with atomic electric dipole transition matrix elements in his famous 1925 paper on quantum mechanics, that is, by deriving commutation relations for the currents and then saturating them by inserting sums over suitable intermediate states [26].  This was the so-called current algebra program.  Among other things, this approach was used by Adler and Weisberger to derive their celebrated formula for the axial-vector coupling constant of beta decay [27].

Sometime around 1965 we began to understand all these developments and how they were related to each other in a more modern way.  It was realized that the strong interactions must have a broken symmetry, $SU(2) \times  SU(2)$, consisting of ordinary isotopic spin transformations plus chiral isotopic spin transformations acting oppositely on the left and right-handed parts of nucleon fields.  Contrary to what I and others had thought at first, such a broken symmetry does not look in the laboratory like an ordinary approximate symmetry.  If it is an exact symmetry, but spontaneously broken, the symmetry implications are found in precise predictions for the low-energy interactions of the massless Goldstone bosons, which for $SU(2) \times SU(2)$ would be the pions.  Among these ``soft pion'' formulas is the Goldberger--Treiman relation, which should be read as a formula for the pion-nucleon coupling at zero pion momentum.  Of course $SU(2) \times SU(2)$ is only an approximate symmetry of the strong interactions, so the pion is not a massless particle, but is what (over Goldstone's objections) I later called a pseudo-Goldstone boson, with an exceptionally small mass.   

From this point of view one can calculate things having nothing to do with the electro-weak interactions, nothing to do with the semi-leptonic vector and axial vector currents, but that refer solely to the strong interactions.  Starting in 1965, the pion-nucleon scattering lengths were calculated independently by Tomozawa and myself [28], and I calculated the pion-pion scattering lengths [29]. Because these processes involve more than one soft pion, the results of these calculations depended critically on the $SU(2) \times SU(2)$ symmetry.  This work had a twofold impact. One is that it tended to kill off the S-matrix approach to the strong interactions, because although there was nothing wrong with the S-matrix philosophy, its practical implementation relied on the pion-pion interaction being rather strong at low energy, while these new results showed that it the interaction is in fact quite weak at low energy.  This work also tended for a while to reduce interest in what Higgs and Brout and Englert had done, for we no longer wanted to get rid of the nasty Goldstone bosons (as had been hoped particularly by Higgs), because now the pion was recognized as a Goldstone boson, or very nearly.

This brings me to the electroweak theory, as developed by myself [30], and independently by Salam [31].  Unfortunately Salam is not with us to describe the chain of reasoning that led him to this theory, so I can only speak about my own work.  My starting point in 1967 was the old aim, going back to Yang and Mills, of developing a gauge theory of the strong interactions, but now based on the symmetry group that underlies the successful soft-pion predictions, the symmetry group $SU(2) \times SU(2)$ [32].  I supposed that the vector gauge boson of this theory would be the $\rho$-meson, which was an old idea, while the axial-vector gauge boson would be the $a_1$ meson, an enhancement in the $\pi-\rho$ channel which was known to be needed to saturate certain spectral function sum rules, which I had developed a little earlier that year [33].   Taking the $SU(2) \times SU(2)$ symmetry to be exact but spontaneously broken, I encountered the same result found earlier by Higgs and Brout and Englert; the Goldstone bosons disappeared and the $a_1$ meson became massive.  But with the isotopic spin subgroup unbroken, then (in accordance with a general result of Kibble [34]) the $\rho$-meson would remain massless.   I could of course put in a common mass for the $a_1$ and $\rho$ by hand, which at first gave encouraging results.  The pion now reappeared as a Goldstone boson, and the spontaneous breaking of the symmetry made the $a_1$ mass larger than the $\rho$ mass by a factor of the square root of two, which was just the ratio that had come out of the spectral function sum rules.  For a while I was encouraged, but the theory was really too ugly.  It was the same old problem: putting in a $\rho$-meson mass or any gauge boson mass by hand destroyed the rationale for the theory and made the theory less predictive, and it also made the theory not renormalizable. So I was very discouraged. 

Then it suddenly occurred to me that this was a perfectly good sort of theory, but I was applying it to the wrong kind of interaction.  The right place to apply these ideas was not to the strong interactions, but to the weak and electromagnetic interactions.  There would be a spontaneously broken gauge symmetry (probably not $SU(2) \times SU(2)$) leading to massive gauge bosons that would have nothing to do with the $a_1$ meson but could rather be identified with the intermediate vector bosons of the weak interactions.   There might be some generator of the gauge group that was not spontaneously broken, and the corresponding massless gauge boson would not be the $\rho$ meson, but the photon.   The gauge symmetry would be exact; there would be no masses put in by hand. 

I needed a concrete model to illustrate these general ideas.  At that time I didn't have any faith in the existence of quarks, and so I decided only to look at the leptons, and somewhat arbitrarily I decided to consider only symmetries that acted on just one generation of leptons, separately from antileptons --- just the left-handed electron and electron-type neutrino, and the right-handed electron.   With those ingredients, the largest gauge group you could possibly have would be $SU(2) \times U(1) \times U(1)$.  One of the $U(1)$s could be taken to be the gauge group of lepton conservation.  Now, I knew that lepton number was conserved to a high degree of accuracy, so this $U(1)$ symmetry was presumably not spontaneously broken, but I also knew that there was no massless gauge boson associated with lepton number, because according to an old argument of Lee and Yang [35] it would produce a force that would compete with gravitation.  So I decided to exclude this part of the gauge group, leaving just $SU(2) \times U(1)$ gauge symmetry.  The gauge bosons were then the charged massive particle (and its antiparticle) that had traditionally been called the $W$; a neutral massive vector particle that I called the $Z$; and the massless photon.  The interactions of these gauge bosons with the leptons and with each other were fixed by the gauge symmetry.  Afterwards I looked back at the literature on intermediate vector boson theories from the late 1950s and early 1960s, and I found that the global $SU(2) \times U(1)$ group structure had already been proposed in 1961 by Glashow [12].  I only learned later of the independent 1964 work of Salam and Ward [12].   I think the reason that the four of us had independently come to the same $SU(2) \times U(1)$ group structure is simply because with these fermionic ingredients, just one generation of leptons, there is no other group you can be led to.   But now the theory was based on an exact though spontaneously broken gauge symmetry. 

The spontaneous breakdown of this symmetry had not only to give mass to the intermediate vector bosons of the weak interactions, it also had to give mass to the electron (and also, in another lepton doublet, to the muon.)   The only scalar particles whose vacuum expectation values could give mass to the electron and the muon would have to form $SU(2) \times U(1)$ doublets with charges +e and zero.  For simplicity,  I assumed that these would be the only kind of scalar fields in the theory.  That made the theory quite predictive. It allowed the masses of the $W$ and the $Z$ as well as their couplings to be calculated in terms of a single unknown angle $\theta$.  Whatever the value of $\theta$, the $W$ and $Z$ masses were predicted to be quite large, large enough to have escaped detection.  The same results apply with several scalar doublets.  (These predictions by the way could also have been obtained in a ``technicolor'' theory in which the electroweak gauge symmetry is spontaneously broken by strong forces, as realized twelve years later by Susskind and myself [36].  This is still a possibility, but such technicolor theories have problems, and I'm betting on the original scalar doublet or doublets.) 

In addition to predicting the masses and interactions of the $W$ and $Z$ in terms of a single angle, the electroweak theory made another striking prediction which could not be verified at the time, and still has not been.  A single scalar doublet of complex scalar fields can be written in terms of four real fields.  Three of the gauge symmetries of $SU(2) \times U(1)$ are spontaneously broken, which eliminates the three Goldstone bosons associated with these fields.  This leaves over one massive neutral scalar particle, as a real particle that can be observed in the laboratory.  This particle, which first made its appearance in the physics literature in 1967 [30], has so far not made its appearance in the laboratory.  Its couplings were already predicted in this paper, but its mass is still unknown.  To distinguish this particle from the Goldstone bosons it has come to be called the Higgs boson, and it is now a major target of experimental effort.  With several doublets (as in supersymmetry theories) there would be several of these particles, some of them charged.

Both Salam and I guessed that the electroweak theory is renormalizable, because we had started with a theory that was manifestly renormalizable. But the theory with spontaneous symmetry breaking had a new perturbative expansion, and the question was whether or not renormalizability was preserved in the new perturbation theory.  We both said that we thought that it was, but didn't prove it. I can't answer for Salam, but I can tell you why I didn't prove it. It was because at that time I disliked the only method by which it could be proved --- the method of path integration.  There are two alternative approaches to quantization: the old operator method that goes back to the 1920s, and Feynman path-integration [37].  When I learned the path-integration approach in graduate school and subsequent reading, it seemed to me to be no more powerful than the operator formalism, but with a lot more hand-waving.  I tried to prove the renormalizability of the electroweak theory using the most convenient gauge that can be introduced in the operator formalism, called unitarity gauge, but I couldn't do it [38].  I suggested the problem to a student [39], but he couldn't do it either, and to this day no one has done it using this gauge.  What I didn't realize was that the path-integral formalism allows the use of gauges that cannot be introduced as a condition on the operators in a quantum field theory, so it gives you a much larger armamentarium of possible gauges in which gauge invariant theories can be formulated. 

Although I didn't understand the potentialities of path integration,  Veltman and his student t'Hooft did.  In 1971 t'Hooft used path integration to define a gauge in which it was obvious that spontaneously broken non-Abelian gauge theories with only the simplest interactions had a property that is essential to renormalizability, that in all orders of perturbation theory there are only a finite number of infinities [40].  This did not quite prove that the theory was renormalizable, because the Lagrangian is constrained by a spontaneously broken but exact gauge symmetry.  In the `t Hooft gauge it was obvious that there were only a finite number of infinities, but how could one be sure that they exactly match the parameters of the original theory as constrained by gauge invariance, so that these infinities can be absorbed into a redefinition of the parameters?  This was initially proved in 1972 by Lee and Zinn-Justin [41] and by 't Hooft and Veltmann [42], and later in an elegant formalism by Becchi, Rouet, and Stora, and by Tyutin [43].  But I must say that after 't Hooft's original 1971 paper, (and, for me, a subsequent related paper by Ben Lee [44]) most theorists were pretty well convinced that the theory was renormalizable, and at least among theorists there was a tremendous upsurge of interest in this kind of theory. 

From today's perspective, it may seem odd that so much attention was focused on the issue of renormalizability.  Like general relativity, the old theory of weak interactions based on four-fermion interactions could have been regarded as an effective quantum field theory [45], which works perfectly well at sufficiently low energy, and with the introduction of a few additional free parameters even allows the calculation of quantum corrections.  The expansion parameter in such theories is the energy divided by some characteristic mass and as long as you work to a given order in the energy you will only need a finite number of coupling types, so that the coupling parameters can absorb all of the infinities.  But such theories inevitably lose all predictive power at energies above the characteristic mass.  For the four-fermion theory of weak interactions it was clear that the characteristic mass was no greater than about 300 GeV, and as we now know, it is actually of the order of the $W$ mass.  The importance of the renormalizability of the electroweak theory was not so much that infinities could be removed by renormalization, but rather that the theory had the potentiality of describing weak and electromagnetic interactions at energies much greater than 300 GeV, and perhaps all the way up to the Planck scale.  The search for a renormalizable theory of weak interactions was the right strategy but, as it turned out, not for the reasons we originally thought. 

These attractive theories of the electroweak theory did not mean that the theory was true --- that was a matter for experiment.  After the demonstration that the electroweak theory is renormalizable, its experimental consequences began to be taken seriously.  The theory predicted the existence of neutral currents, but this was an old story.  Suggestions of neutral weak currents can be traced back to 1937 papers of Gamow and Teller, Kemmer, and Wentzel [46].  Neutral currents had appeared in the 1958 paper by Bludman and in all the subsequent papers in ref. 12, including of course those of Glashow and of Salam and Ward.  But now there was some idea about their strength.  In 1972 I looked at the question of how easy it would be to find semi-leptonic neutral current processes, and I found that although in the electroweak theory they are somewhat weak compared to the ordinary charged-current weak interactions, they were not too weak to be seen [47].  In particular, I pointed out that the ratio of elastic neutrino-proton scattering to the corresponding inelastic charged-current reaction would have a value between .15 and .25, depending on the value of the unknown angle $\theta$.  A 1970 experiment [48] had given a value of .12 plus or minus .06 for this ratio, but the experimenters didn't believe that they were actually seeing neutral currents, so they didn't claim to have observed a neutral current reaction at a level of roughly 12\% of the charged current reaction, and instead quoted this result as an upper bound.  The minimum theoretical value 0.15 of this ratio applies for $\sin^2\theta= 0.25$, which is not far from what we now know is the correct value.  I suspect that this 1970 experiment had actually observed neutral currents, but you get credit for making discoveries only when you claim that you have made the discovery.
 
Neutral currents were discovered in 1973 at CERN [49].  I suspect that this will be mentioned later today, so I won't go into it here.  At first the data on neutral current reactions looked like it exactly fit the electroweak theory, but then a series of other experiments gave contrary results. The most severe challenge came in 1976 from two atomic physics experiments [50] that seemed to show that there was no parity violation in the bismuth atom at the level that would be expected to be produced by neutral current electron-nucleon interactions in the electroweak theory.  For most theorists these experiments did not challenge the basic idea that weak interactions arise from a spontaneously broken gauge symmetry, but they threw serious doubt on the specific $SU(2) \times U(1)$ implementation of the idea. Many other models were tried during this period, all sharing the property of being terribly ugly. Finally, parity violation in the neutral currents was discovered at the expected level in 
electron--nucleon scattering at SLAC in 1978 [51], and after that most physicists took it for granted that the electroweak theory is essentially correct. 

The other half of the Standard Model is quantum chromodynamics.  By the early 1970s the success of the electroweak theory had restored interest in Yang--Mills theory.  In 1973 Gross and Wilczek and Politzer independently discovered that non-Abelian gauge theories have the remarkable property of asymptotic freedom [52].  They used renormalization group methods due to Gell-Mann and Low [53], which had been revived in 1970 by Callan, Symanzik, Coleman and Jackiw [54], to define an effective gauge coupling constant as a function of energy, and they showed that in Yang--Mills theories with not too many fermions this coupling goes to zero as the energy goes to infinity.  (`t Hooft had found this result and announced it at a conference in 1972, but he waited to publish this result and work out its implications while he was doing other things, so his result did not attract much attention.)  It was already known both from baryon systematics and from the rate of neutral pion decay into two photons that quarks of each flavor u, d, s, etc. must come in three colors [55], so it was natural to take the gauge symmetry of the strong interactions as an $SU(3)$ gauge group acting on the three-valued color quantum number of the quarks.  Subsequent work [56] by Gross and Wilczek and by Georgi and Politzer using the Wilson operator product expansion [57] showed that the decrease of the strong coupling constant with increasing energy in this theory explained why ``partons'' had appeared to be weakly coupled in the 1968 Friedman--Kendall--Taylor experiment [7].

But a big problem remained: what is one to do with the massless $SU(3)$ gauge bosons, the gluons? The original papers [52] of Politzer and Gross and Wilczek suggested that the reason why massless gluons are not observed is that the gauge symmetry is spontaneously broken, just as in the electroweak theory.  The gluons could then be assumed to be too heavy to observe. Very soon afterwards a number of authors independently suggested an alternative, that the gauge symmetry is not broken at all, the gluons are in fact massless, but we don't see them for the same reason that we don't see the quarks, which is that, as a result of the peculiar infrared properties of non-Abelian gauge theories, color is trapped; color particles like quarks and gluons can never be isolated [58].  This has never been proved.  There is now a million dollar prize offered by the Cray Foundation to anyone who succeeds in proving it rigorously, but since it is true I for one am happy to leave the proof to the mathematicians.  

One of the great things that came out of this period of the development of the electroweak and the strong interaction theories is an understanding at long last of the old approximate symmetries. It was now understood that these symmetries were approximate because they weren't fundamental symmetries at all; they were just accidents.  Renormalizable quantum chromodynamics must respect strangeness conservation and charge conjugation invariance, and, aside from a non-perturbative effect that I don't have time to go into, it must also respect parity and time reversal invariance. You cannot introduce any renormalizable interaction into the theory that would violate those symmetries.  This would not be true if scalar fields participated in the strong interactions, as in the old Yukawa theory.  This result was not only aesthetically pleasing, but crucial, because if there were possible renormalizable interactions that violated, say, strangeness conservation, or parity, then even if you didn't put such interactions in the theory, higher order weak interactions would generate them at first order in the fine structure constant [59].  There would then be violations of parity and strangeness conservation in the strong interactions at a level of a percent or so, which certainly is not the case.  

If one makes the additional assumption that the up, down and strange quark masses are small, then without having to assume anything about their ratios it follows that the theory has an approximate $SU(3) \times SU(3)$ symmetry, including not only the eightfold way but also the spontaneously broken chiral $SU(2) \times SU(2)$ symmetry that had been used to derive theorems for low-energy pions back in the mid 1960s.  Furthermore, with an intrinsic $SU(3) \times SU(3)$ symmetry breaking due to small up, down and strange quark masses, this symmetry gives rise to the Gell-Mann--Okubo mass formula [60] and justifies the symmetry-breaking assumptions made in the 1965 derivation of the pion-pion scattering lengths [29]. Finally, it is automatic in such theories that the semi-leptonic currents of the weak interactions must be symmetry currents associated with this $SU(3) \times SU(3)$ symmetry.  This was a really joyous moment for theorists.  Suddenly, after all those years of dealing with approximate symmetries, it all fell into place.  They are not fundamental symmetries of nature at all; they are just accidents dictated by the renormalizability of quantum chromodynamics and the gauge origin of the electroweak interactions. 

Before closing, I must also say something about two other topics: the problem of strangeness nonconservation in the weak interactions, and the discoveries of the third generation of quarks and leptons and of the $W$ and $Z$.  

The charge exchange semileptonic interactions were long known to violate strangeness conservation, so any charged $W$ boson would have to have couplings in which strangeness changes by one unit.  It follows that the exchange of pairs of $W$s could produce processes like $K-\bar{K}$ conversion in which strangeness changes by two units.  With an ultraviolet cut-off of the order of the $W$ mass, the amplitude for such processes would be suppressed by only two factors of the inverse $W$ mass, like a first-order weak interaction, in contradiction with the known magnitude of the mass difference of the $K_1$ and $K_2$.  A way out of this difficulty was discovered in 1970 by Glashow, Iliopoulos and Maiani [61].  They found that these strangeness-violating first-order weak interactions would disappear if there were two full doublets of quarks, entering in the same way in the weak interactions.  This required a fourth quark, called the charm quark.  They also showed that with the fourth quark in the theory, in an $SU(2)$ gauge theory the neutral currents would not violate strangeness conservation.  In 1972 I showed that the GIM mechanism also works for the $Z$ exchange of the $SU(2) \times U(1)$ electroweak theory [62].  The introduction of the fourth quark also had the happy consequence, as shown independently by Bouchiat, Iliopoulos, and Meyer and by myself [63], that the triangle anomalies that would otherwise make the theory not really gauge invariant all cancelled.  The $K_1 - K_2$ mass difference was calculated as a function of the charm quark mass by Gaillard and Lee [64], who used the experimental value of this mass difference to estimate that the mass of the charm quark would be about 1.5 GeV.  Further, using the new insight from quantum chromodynamics that the strong coupling is not so strong at energies of this order, Applequist and Politzer in 1974 (just before the discovery of the J/psi) predicted that the charm-anticharm bound state would be rather narrow [65].  This narrow bound state was discovered in 1974 [66], and immediately not only provided evidence for the existence of a fourth quark, but also gave vivid testimony that quarks are real. 

The only thing remaining in the completion of the Standard Model was the discovery of the third generation: the $\tau$ lepton [67] (and the corresponding neutrino) and the bottom [68] and top [69] quarks.  This provided a new mechanism for CP violation, the complex phase factor in the 
Cabibbo--Kobayashi--Maskawa matrix [70] appearing in the semi-leptonic weak interactions.  The fact that the third generation of quarks is only slightly mixed in this matrix with the first and second generations even makes it natural that the CP violation produced in this way should be rather weak.  Unfortunately, the explanation of the masses and mixing angles in the Cabibbo--Kobayashi--Maskawa matrix continues to elude us.

These developments were crowned in 1983 with the discovery [71] of the $W$ and the $Z$ intermediate vector bosons.  It has proved possible to measure their masses with great precision, which has allowed a stringent comparison of the electroweak theory with experiment.  This comparison has even begun to give hints of the properties of the as yet undiscovered scalar particle or particles. 

Well, those were great days.  The 1960s and 1970s were a time when experimentalists and theorists were really interested in what each other had to say, and made great discoveries through their mutual interchange.  We have not seen such great days in elementary particle physics since that time, but I expect that we will see good times return again in a few years, with the beginning of a new generation of experiments at this laboratory.  

\begin{center}
{\bf References}
\end{center}

\begin{enumerate}
\item G. Breit, E. U. Condon, and R. D. Present, Phys. Rev. 50, 825 (1936); B. Cassen and E. U. Condon, Phys. Rev. 50, 846 (1936); G. Breit and E. Feenberg, Phys. Rev. 50, 850 (1936).  This symmetry was suggested by the discovery of the equality of proton-proton and proton-neutron forces by M. A. Tuve, N. Heydenberg, and L. R. Hafstad, Phys. Rev. 50, 806 (1936).  Heisenberg had earlier used an isotopic spin formalism, but without introducing any symmetry beyond invariance under interchange of protons and neutrons.

\item M. Gell-Mann, Phys. Rev. 92, 833 (1953); T. Nakano and K. Nishijima, Prog. Theor. Phys. 10, 581 (1955).

\item T. Lee and C. N. Yang, Phys. Rev. 104, 254 (1956); C. S. Wu {\em et al.} Phys. Rev. 105, 1413 (1957); R. Garwin, M. Lederman, and M. Weinrich, Phys. Rev. 105, 1415 (1957); J. I. Friedman and V. L. Telegdi, Phys. Rev. 105, 1681.

\item J. H. Christensen, J. W. Cronin, V. L. Fitch, and R. Turlay, Phys. Rev. Lett. 13, 138 (1964).

\item M. Gell-Mann, Cal. Tech. Synchotron Lab Report CTSL-20 (1961); Y. Ne'eman, Nucl. Phys. 26, 222 (1961).

\item M. Gell-Mann, Phys. Lett. 8, 214 (1964); G. Zweig, CERN preprint TH401 (1964). Earlier, it had been suggested that baryon number should be included in the hadron symmetry group by expanding $SU(3)$ to $U(3)$ rather than $SU(3) \times U(1)$, with each lower or upper index in a tensor representation of $U(3)$ carrying a baryon number $1/3$ or $-1/3$, respectively, by H. Goldberg and Y. Ne'eman, Nuovo Cimento 27, 1 (1963).

\item E. D. Bloom {\em et al.}, Phys. Rev. Lett. 23, 930 (1969); M. Briedenbach {\em et al.}, Phys. Rev. Lett. 23, 935 (1969); J. L. Friedman and H. W. Kendall, Annual Reviews of Nuclear Science 22, 203 (1972).

\item J. D. Bjorken, Phys. Rev. 179, 1547 (1969); R. P. Feynman, Phys. Rev. Lett. 23, 1415 (1969).

\item C. N. Yang and R. L. Mills, Phys. Rev. 96, 191 (1954).

\item B. de Witt, Phys. Rev. Lett. 12, 742 (1964); Phys. Rev. 162, 1195 (1967); L. D. Faddeev and V. N. Popov, Phys. Lett. B 25, 29 (1967); also see R. P. Feynman, Acta Phys. Pol. 24, 697 (1963); S. Mandelstam, Phys. Rev. 175, 1580, 1604 (1968).

\item E. C. G. Sudarshan and R. E. Marshak, in {\em Proceedings of the Padua--Venice Conference on Mesons and Recently Discovered Particles}, p. v-14 (1957); Phys. Rev. 109, 1860 (1958); R. P. Feynman and M. Gell-Mann, Phys. Rev. 109, 193 (1958).

\item J. Schwinger, Ann. Phys. 2, 407 (1957); T. D. Lee and C. N. Yang, Phys. Rev. 108, 1611 (1957); 119, 1410 (1960); S. Bludman, Nuovo Cimento 9, 433 (1958); J. Leite-Lopes, Nucl. Phys. 8. 234 (1958); S. L. Glashow, Nucl. Phys. 22, 519 (1961); A. Salam and J. C. Ward, Phys. Lett. 13, 168 (1964).

\item A. Komar and A. Salam, Nucl. Phys. 21, 624 (1960); H. Umezawa and S. Kamefuchi, Nucl. Phys. 23, 399 (1961); S. Kamefuchi, L. O' Raifeartaigh, and A. Salam, Nucl. Phys. 28, 529 (1961); A. Salam, Phys. Rev. 127, 331 (1962); M. Veltman, Nucl. Phys. B 7, 637 (1968); Nucl. Phys. 21, 288 (1970); D. Boulware, Ann. Phys. 56, 140 (1970).

\item W. Heisenberg, lecture ``What is an Elementary Particle?'' to the German Physical Society on March 5, 1975, reprinted in English translation in {\em Encounters with Einstein And Other Essays of People, Places, and Particles} (Princeton University Press, 1983).

\item J. Goldstone, Nuovo Cimento 19, 154 (1961).

\item J. Goldstone, A. Salam, and S. Weinberg, Phys. Rev. 127, 965 (1962).

\item P. W. Higgs, Phys. Lett. 12, 132 (1964); Phys. Lett. 13, 508 (1964); Phys. Rev. 145, 1156 (1966).  Also see G. S. Guralnik, C. Hagen, and T. W. B. Kibble, Phys. Rev. Lett. 13, 585 (1964).

\item F. Englert and R. Brout, Phys. Rev. Lett. 13, 321 (1964).

\item P. M. Anderson, Phys. Rev. 130, 439 (1963).

\item M. L. Goldberger and S. B. Treiman, Phys. Rev. 111, 354 (1958).

\item M. Gell-Mann and M. L\'{e}vy, Nuovo Cimento 16, 705 (1960); J. Bernstein, S. Fubini, 
M. Gell-Mann, and W. Thirring, Nuovo Cimento 17, 757 (1960); K-C. Chou, Soviet Physics JETP 12, 492 (1961).

\item Y. Nambu, Phys. Rev. Lett. 4, 380 (1960).

\item Y. Nambu and G. Jona-Lasinio, Phys. Rev. 122, 345 (1961).

\item Y. Nambu and D. Lurie, Phys. Rev. 125, 1429 (1962); Y. Nambu and E. Shrauner, Phys. Rev. 128, 862 (1962).  These predictions were generalized by S. Weinberg, Phys. Rev. Lett. 16, 879 (1966).

\item J. Bardeen, L. N. Cooper, and J. R. Schrieffer, Phys. Rev. 108, 1175 (1957).

\item M. Gell-Mann, Physics 1, 63 (1964).

\item S. L. Adler, Phys. Rev. Lett. 14, 1051 (1965); Phys. Rev. 140, B736 (1965); W. I. Weisberger, Phys. Rev. Lett. 14, 1047 (1965); Phys. Rev. 143, 1302 (1965).

\item S. Weinberg, Phys. Rev. Lett. 17, 616 (1966); Y. Tomozawa, Nuovo Cimento 46A, 707 (1066).

\item S. Weinberg, ref. 28.

\item S. Weinberg, Phys. Rev. Lett. 19, 1264 (1967).

\item A. Salam, in {\em Elementary Particle Physics}, N. Svartholm, ed. (Nobel Symposium No. 8, Almqvist \& Wiksell, Stockholm, 1968), p. 367.

\item This work was briefly reported in ref. 33, footnote 7.

\item S. Weinberg, Phys. Rev. Lett. 18, 507 (1967).

\item T. W. B. Kibble, Phys. Rev. 155, 1554 (1967).

\item T. D. Lee and C. N. Yang, Phys. Rev. 98, 101 (1955).

\item S. Weinberg, Phys. Rev. D19, 1277 (1979); L. Susskind, Phys. Rev. D19, 2619 (1979).

\item R. P. Feynman, ``The Principle of Least Action in Quantum Mechanics'' (Princeton University Ph. D. thesis, 1942; University Microfilms Publication No. 2948, Ann Arbor.)  This work was in the context of non-relativistic quantum mechanics.  Feynman later applied this formalism to the Dirac theory of electrons, but its application to a full-fledged quantum field theory was the work of other authors, including some of those in ref. 10.  

\item I reported this work later in Phys. Rev. Lett. 27, 1688 (1971) and described it in more detail in Phys. Rev. D 7, 1068 (1973).

\item See L. Stuller, M.I.T. Ph.D. thesis (1971).

\item G. `t Hooft, Nucl. Phys. B 35, 167 (1971).

\item B. W. Lee and J. Zinn-Justin, Phys. Rev. D 5, 3121, 3137, 3155 (1972).

\item G. `t Hooft and M. Veltman, Nucl. Phys. B 44, 189 (1972); Nucl. Phys. B 50. 318 (1972).

\item C. Becchi, A. Rouet, and R. Stora, Commun. Math. Phys. 42, 127 (1975); Ann. Phys. 98, 287 (1976); I. V. Tyutin, Lebedev Institute preprint N39 (1975).

\item B. W. Lee, Phys. Rev. D 5, 823 (1972). 

\item S. Weinberg, Physica 96A, 327 (1979).

\item G. Gamow and E. Teller, Phys. Rev. 51, 289L (1937); N.Kemmer, Phys. Rev. 52, 906 (1937); G. Wentzel, Helv. Phys. Acta 10, 108 (1937).

\item S. Weinberg, Phys. Rev. 5, 1412 (1972).

\item D. C. Cundy {\em et al.}, Phys. Lett. B 31, 478 (1970).

\item F. J. Hasert {\em et al.}, Phys. Lett. B 46, 121, 138 (1973); P. Musset {\em et al.}, J. Phys. (Paris) 11/12, T34 (1973).

\item L. L. Lewis {\em et al.}, Phys. Rev. Lett. 39, 795 (1977); P. E. G. Baird {\em et al.}, Phys. Rev. Lett. 39, 798 (1977).

\item C. Y. Prescott {\em et al.}, Phys. Lett. 77B, 347 (1978).

\item D. J. Gross and F. Wilczek, Phys. Rev. Lett. 30, 1343 (1973); H. D. Politzer, Phys. Rev. Lett. 30, 1346 (1973).  

\item M. Gell-Mann and F. E. Low, Phys. Rev. 95, 1300 (1954).

\item C. G. Callan, Phys. Rev. D2, 1541 (1970); K. Symanzik, Commun. Math. Phys. 18, 227 (1970); C. G. Callan, S. Coleman, and R. Jackiw, Ann. of Phys. (New York) 47, 773 (1973). 

\item O. W. Greenberg, Phys. Rev. Lett. 13, 598 (1964); M. Y. Han and Y. Nambu, Phys. Rev. B 139, 1006 (1965); W. A. Bardeen, H. Fritzsch, and M. Gell-Mann, in {\em Scale and Conformal Symmetry in Hadron Physics}, R. Gatto, ed. (Wiley, New York, 1973), p. 139.

\item H. Georgi and H. D. Politzer, Phys. Rev. D 9, 416 (1974); D. J. Gross and F. Wilczek, Phys. Rev. D 9, 980 (1974).

\item K. Wilson, Phys. Rev. 179, 1499 (1969).

\item S. Weinberg, Phys. Rev. Lett. 31, 494 (1973); D. J. Gross and F. Wilczek, Phys. Rev. D 8, 3633 (1973); H. Fritzsch, M. Gell-Mann, and H. Leutwyler, Phys. Lett. B 47, 365 (1973).

\item S. Weinberg, ref. 58.

\item M. Gell-Mann, ref. 5; S. Okubo, Prog. Theor. Phys. 27, 949 (1962).

\item S. Glashow, J. Iliopoulos, and L. Maiani, Phys. Rev. D 2, 1285 (1970).

\item S. Weinberg, ref. 47.

\item C. Bouchiat, J. Iliopoulos, and P. Meyer, Phys. Lett. 38B, 519 (1972); S. Weinberg, in {\em Fundamental Interactions in Physics and Astrophysics}, eds. G. Iverson {\em et al.} (Plenum Press, New York, 1973), p. 157.

\item M. Gaillard and B. W. Lee, Phys. Rev. D 10, 897 (1974).

\item T. Appelquist and H. D. Politzer, Phys. Rev. Lett. 34, 43 (1975).

\item J. J. Aubert {\em et al.}, Phys. Rev. Lett. 33, 1404 (1974); J. E. Augustin {\em et al.}, Phys. Rev. Lett. 33, 1406 (1974).

\item M. Perl {\em et al.} Phys. Rev. Lett. 35, 195, 1489 (1975); Phys. Lett. 63B, 466 (1976).

\item S. W. Herb {\em et al.}, Phys. Rev. Lett. 39, 252 (1975).

\item F. Abe {\em et al.}, Phys. Rev. Lett. 74, 2626 (1995); S. Abachi {\em et al.}, Phys Rev. Lett. 74, 2632 (1995).

\item N. Cabibbo, Phys. Rev. Lett. 10, 531 (1963); M. Kobayashi and K. Maskawa, Prog. Theor. Phys. 49, 282 (1972).

\item G. Arnison {\em et al.}, Phys. Lett. 122B, 103 (1983); 126B, 398 (1983); 129B, 273 (1983); 134B, 469 (1984); 147B, 241 (1984).

\end{enumerate}
\end{document}